 \documentclass[12pt,preprint]{aastex}

\slugcomment{}

\shorttitle{Intergalactic Globular Clusters in Coma}
\shortauthors{Mar\' \i n-Franch, A.} 
\begin{document}

\title{Intergalactic globular clusters and the faint end of the galaxy number counts in A1656 (Coma)}

\author{A. Mar\' \i n-Franch}\affil{Instituto de Astrof\' \i sica de Canarias, E-38205 La Laguna,
Tenerife, Spain}
\email{amarin@ll.iac.es}

\and

\author{A. Aparicio}
\affil{Instituto de Astrof\' \i sica de Canarias, E-38205 La Laguna,
Tenerife, Spain}
\email{aaj@ll.iac.es}

\begin{abstract}

The existence of an intergalactic globular cluster population in the Coma
cluster of galaxies has been tested using surface-brightness
fluctuations. The main result is that the intergalactic globular cluster
surface density ($N_{\rm IGC}$) does not correlate with the distance to the center
of Coma and hence with the environment. Furthermore, comparing these results
with different Coma mass-distribution model predictions, it is suggested that
$N_{\rm IGC}$ must in fact be zero all over Coma. On the other hand, the results
for $N_{\rm IGC}$ and the faint end of the galaxy number counts (beyond
$m_R=23.5$) are connected. So $N_{\rm IGC}=0$ settles  the slope of this
function, which turns out to be $\gamma=0.36\pm0.01$ down to $m_R=26.5$.

The fact that $N_{\rm IGC}=0$ all over Coma suggests that globular clusters were formed only, or almost only, from protogalactic clouds. None, or
perhaps very few, could have formed in isolated regions. It also seems  inappropriate to advocate a relationship between
intergalactic globular clusters and dark matter distributions, although it is
true that the relationship could still exist but not be strong enough to
have been detected. Finally, since our conclusion is that intergalactic
globular clusters do not exist in Coma, accretion of intergalactic globular
clusters might not be  significant  in galaxy formation and evolutionary
processes in the Coma galaxies.

\end{abstract}

\section{Introduction}\label{intro}

Intergalactic globular clusters (IGCs) are not bound to any particular galaxy but move freely in the potential wells of galaxy clusters. Table 1 will help the reader with the notation used in this paper. The existence of IGCs has been considered by a number of authors. The first discussion about them was by \citet{B56}, who measured the distance to the Abell No.~4 globular cluster (GC). Obtaining a distance modulus of 20.8, he  concluded that it is therefore an {\it intergalactic tramp}. Van den Bergh (1958) estimated that the number of IGCs within the Local Group is one third of the total number of GCs. But the first real quantitative studies of IGCs in galaxy clusters were done by \citet{F82} and \citet{M87}. 

\citet{M87} suggested that galaxies in galaxy clusters suffer several dynamical effects 
owing to encounters with  neighboring galaxies and  to the action of the general galaxy cluster field;  as a result, galaxies lose some GCs. Since these processes are very sensitive to the distribution of the total mass of the galaxy cluster, the study of  IGCs may help us to solve the problem of how the {\it missing mass} is distributed. In this scenario, the number density of lost IGCs should follow the total mass distribution; hence, they might be concentrated toward the center of the galaxy cluster. In the same context, \citet{W87} suggested that cD envelopes and the diffuse light in Coma have the same origin: they are composed of stars tidally stripped from galaxies during  Coma's collapse. So GCs might have been stripped as well, creating an IGC population. 

Alternatively, if IGCs do exist in galaxy clusters, they may have been formed in situ in scenarios such as biased GC formation. \citet{W93} argued that the number density of GCs in biased formation scenarios is extremely sensitive to the presence and amplitude of a background field, such as galaxy cluster-sized and galaxy-sized perturbations in the primordial matter distribution. GCs would have been much more likely to form in developing protogalaxies than in complete isolation, but a small number of IGCs is expected. In these scenarios, the number density of IGCs would be very small, and also might follow the galaxy cluster mass distribution.

From a galaxy formation point of view, as GCs are thought to be among the oldest objects in the Universe, they provide useful information about the galaxy formation and evolution processes. By studying a globular cluster system (GCS) associated with a single galaxy, the history of the host galaxy is probed. The hypothesis of the existence of an IGC population has been used to gain insight into the high-$S_N$ problem. \citet{W95} argued that a high value of $S_N$ in galaxies located near the centers of galaxy clusters is the result of the accretion of a number of IGCs. They assumed that  an IGC population exists in all galaxy clusters, and that the number density of IGCs is concentrated toward the center of the galaxy cluster.

Recently, \citet{C01} did a dynamical analysis of the GCS associated with M87. They concluded that the metal-rich GCs of M87 formed with the galaxy, and 
that the metal-poor GCs are gradually accreting onto M87 along Virgo's principal axis. This could therefore be an example of how the accretion of IGCs would influence the galaxy's evolutionary history.

Although some authors consider that the accretion of IGCs is capable of modifying the properties of the GCS of a galaxy, there are a number of others that defend the view that the greater part of a GCS is formed {in situ} from the protogalactic cloud \citep{B97,F97,H98,M99}. So the possible influence of the accretion of IGCs in galaxy formation and evolutionary models is still an open issue. In this context, the existence or not of IGCs and, if they do exist, their abundance and distribution become crucial ingredients in the development of a unified galaxy evolution scenario.

Even though the existence of IGCs is often assumed, there is little evidence to date to prove this. Do GCs form only during starbursts in galaxies or can they form in isolation? If IGCs exist in all galaxy clusters, their number and spatial distribution are fundamental pieces of information for testing the different GC formation models. Furthermore, the influence of IGCs in the evolution of elliptical galaxies is still not clear. Is accretion of IGCs  important  during the galaxy's formation and evolutionary history? If this is  the case, is it important for all elliptical galaxies, or only for those located in some privileged locations inside galaxy clusters? In this paper the existence of an IGC population in the Coma galaxy cluster has been tested and quantified using the surface-brightness fluctuations (SBF) technique. 
 
Our purpose is to obtain information about the IGC population from the study
of the unresolved point sources by means of SBF analysis. The measured SBF
signal is  produced mainly by faint galaxies and the IGC population (if such a
thing  exists). For this reason, during SBF measurements, a good
characterization of the faint end of the background differential galaxy number counts, $n(m)$, is fundamental. A number of authors  have measured the slope of the faint end of $n(m)$ in the $R$ filter. Results range from $\gamma=0.39$ \citep{T88} to $\gamma$ = 0.31--34 \citep{SH93}. This slope has also been measured in the Hubble Deep Field \citep{W96}, with $\gamma=0.36$ for the $V$ filter and $\gamma=0.31$ for the $I$ filter down to magnitude 26. For the intermediate $R$ filter, the slope must be  between $\gamma=0.31$ and 0.36. From the SBF analysis presented in this paper, we provide not only information about the existence or not of an IGC population, but also an estimate of the slope of the faint end of $n(m)$. 

\section{Observations and data reduction}
 
Observations of the 16 giant elliptical galaxies and the 4 ``blank" regions studied in this paper were done on 2000 April 25 and 27, with the 2.5 m 
Isaac Newton Telescope (INT) at the Roque de los Muchachos Observatory (La
 Palma), using the Wide Field Camera and the Sloan $R$ Filter. This work is based on the results obtained by \citet{MA02}, where the observations, photometric calibration and data reduction procedures are described in detail, as well as the SBF analysis. We provide here only a short outline of the latter.

The concept of SBF was introduced by \citet{TS88}, who noted that, in the surface photometry of a galaxy too far to be resolved, a pixel-to-pixel fluctuation is observed due to the Poisson statistics of the spatial distribution of stars, GCs, background galaxies, etc. The variance of the fluctuation depends on the stellar population, the GCS, background galaxies, foreground stars, and, of course, the distance. It can be assumed that the total pixel-to-pixel variance of an image is the sum of all the independent contributions.

The SBF technique involves spectral analysis of the signal and provides
as a result the total PSF-convolved variance ($P_0$), produced by all objects whose spatial flux distribution is convolved with the PSF:
\begin{equation}
P_0=\sigma^{2}_{\rm sp}+\sigma^{2}_{\rm GC}+\sigma^{2}_{\rm BG}+\sigma^{2}_{\rm fs}, \end{equation}
where $\sigma^{2}_{\rm sp}$, $\sigma^{2}_{\rm GC}$, $\sigma^{2}_{\rm BG}$, and $\sigma^{2}_{\rm fs}$ are the variances produced by the stellar populations, GCs, background galaxies, and foreground stars, respectively. If a population of IGCs is being analyzed, $\sigma^{2}_{\rm sp}$ can be neglected, and hence 
\begin{equation}\label{p0}
P_0=\sigma^{2}_{\rm IGC}+\sigma^{2}_{\rm BG}+\sigma^{2}_{\rm fs}, 
\end{equation}
where $\sigma_{\rm IGC}^2$ is the pixel-to-pixel variance produced by the IGC population where this exists. 

The pixel-to-pixel variance produced by a class of object can be evaluated as the second moment of the differential number counts of that object population. In this way, $\sigma^{2}_{\rm BG}$ can be obtained from the background $n(m)$, fitted to the resolved point sources, and $\sigma^{2}_{\rm fs}$ from the predicted star counts, computed making use of a Galaxy model. 

Once $\sigma^{2}_{\rm BG}$ and $\sigma^{2}_{\rm fs}$ are estimated and $P_0$ is measured, $\sigma^{2}_{\rm IGC}$ can be easily obtained from equation~\ref{p0}. Note that in Coma, GCs cannot be directly resolved from ground-based observations (except the brightest individuals), so, in order to determine the surface number density of IGCs, SBF analysis becomes necessary. 

The surface number density of IGCs ($N_{\rm IGC}$) can be estimated from $\sigma^{2}_{\rm IGC}$ assuming a fixed GC luminosity function ($\varphi$). A fixed $\varphi$ provides a fixed $\sigma^{2}_{\rm IGC}/N_{\rm IGC}$ ratio that can be used to transform variances into surface densities. This ratio depends on the photometric calibration.

Mar\'\i n-Franch \& Aparicio's (2002) results used here are listed in Table \ref{igc}. The name of the region studied  is listed in column 1. In column 2, the distance to the Coma galaxy cluster center is also shown. Finally, the obtained $P_0$ in the outer region of each galaxy (the
region further away from the galactic center, where galactic GCs no longer exist) and in the ``blank" areas are listed in column 3. The following discussion is based on these results.

\section{Results} \label{results}

Let us consider regions close to the studied galaxies, but beyond the galactic
halos. If IGCs  existed then $\sigma_{\rm IGC}^2$ would be
\begin{equation}\label{sigmaIGC}
\sigma_{\rm IGC}^2=P_0-\sigma_{\rm BG}^{2}-\sigma_{\rm fs}^{2}.
\end{equation}
$P_0$ is measured from SBF analysis. But, in order to obtain $N_{\rm IGC}$, $\sigma_{\rm BG}^{2}$ and $\sigma_{\rm fs}^{2}$ must be deduced first.  

\subsection{Contribution of foreground stars}

Coma is in a nearly perpendicular direction to the Galactic plane, so the
expected number of foreground stars is very low and its contribution to the
background variance is negligible when compared with the variance produced by
the GCSs of the Coma galaxies \citep{MA02}. But foreground stars can produce
a significant contribution to the total variance in intergalactic
regions, so they must be considered if the variance produced by IGCs and the faint end of $n(m)$ is to be analyzed. 

\begin{figure} 
\plotone{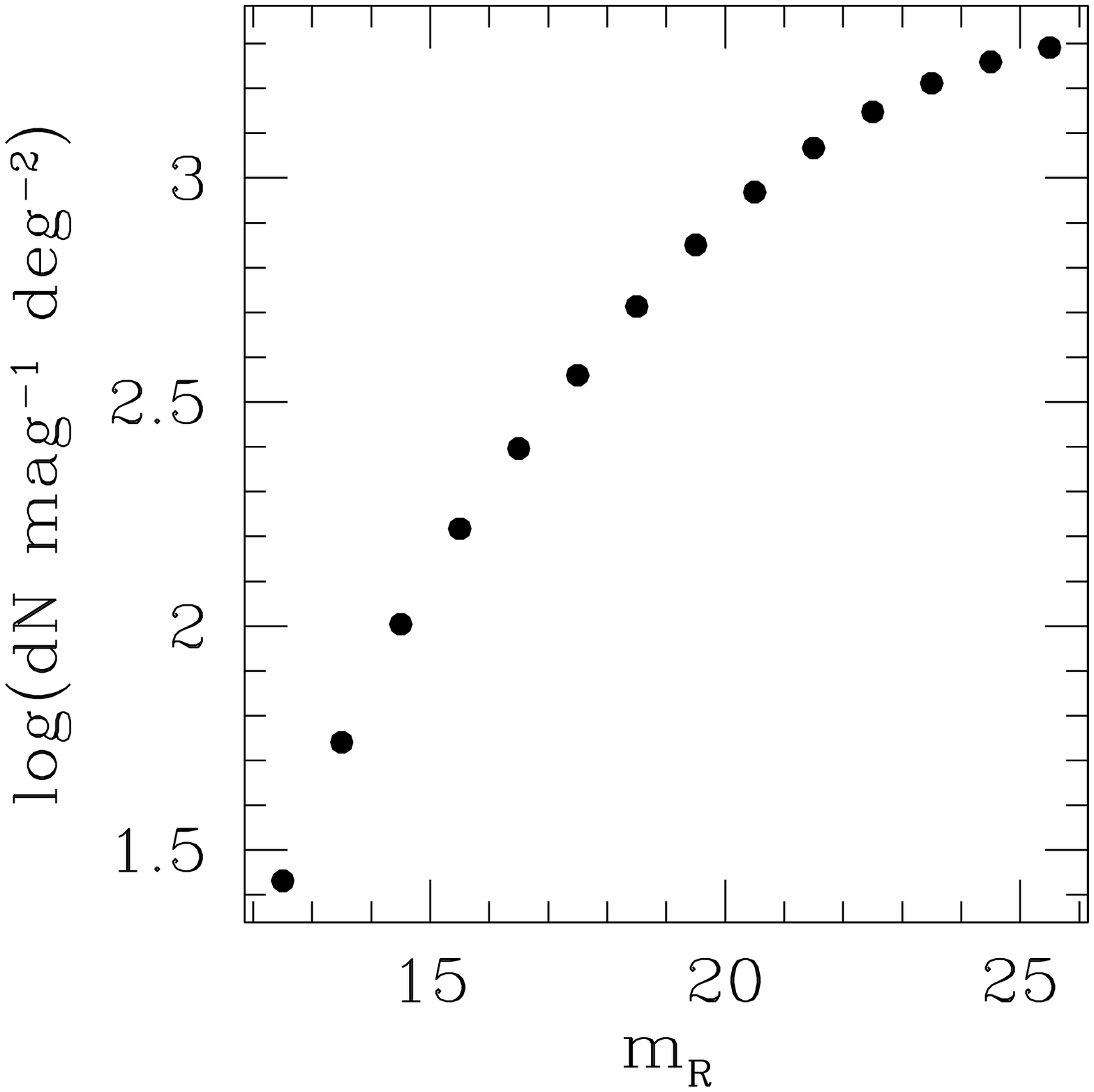}
\caption{Star counts predicted by the Bahcall-Soneira Galaxy model in the direction of the North Galactic Pole \citep{BS81}. \label{stars}}

\end{figure}

\citet{BS81} computed the predicted star counts in selected fields and
photometric bands using their model for the Galaxy. In particular,
they computed the predicted star counts in the northern Galactic pole in the
$R$ filter down to magnitude 26.5 (Fig. \ref{stars}). By computing the second moment of these star counts, $\sigma_{\rm fs}^2$ can be obtained for the 
North Galactic Pole. The result is 
\begin{equation}
\sigma_{\rm fs}^2=(1.50 \pm 0.15)\times10^{-4} \left(\frac{e^-}{\rm s \times pix}\right)^2.
\end{equation}
assuming a $10\%$ uncertainty for the star counts model predictions. As Coma is very close to the North Galactic Pole, we adopt this value and
assume that it is constant over Coma.

\subsection{Faint end of the differential galaxy number counts, $n(m)$}

The background galaxy variance, $\sigma_{\rm BG}^2$, has to be obtained from
the differential galaxy number counts, $n(m)$. \citet{T88} obtained a slope of
$\gamma=0.39$ for the faint end of $n(m)$ in filter $R$. More recently,
\citet{SH93} obtained slopes in the range $\gamma$ = 0.30--0.34, also in
$R$. Unfortunately, the adopted value of $\gamma$ has considerable effects
in the resulting $N_{\rm IGC}$. For this reason, we have evaluated $N_{\rm IGC}$
considering three values for the $n(m)$ slope, $\gamma=0.30$, $\gamma=0.34$,
and $\gamma=0.39$, and has  analyzed the results separately. The corresponding zero
points of $n(m)$ have been obtained from fits to the resolved point sources
measured by \citet{MA02} in which the slope is a fixed parameter.

\begin{figure} 
\plotone{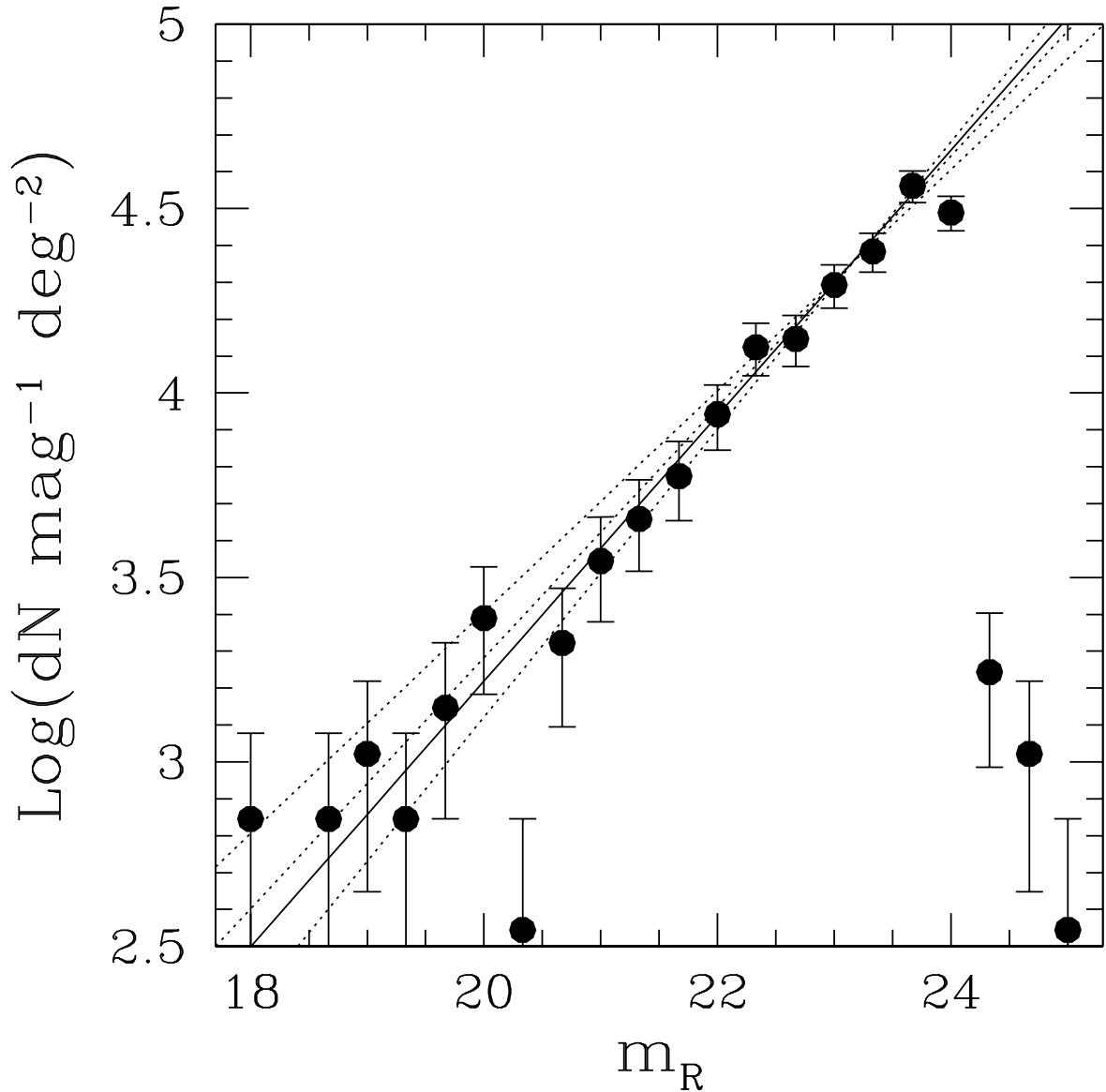}
\caption{Background differential galaxy number counts, $n(m)$, fitted to resolved
point source distribution in the region ``Blank 3". The fit has been
performed using the three values for the $n(m)$ slope $\gamma=0.30$, $0.34$,
and $0.39$ discussed in the text as fixed parameters. The error bars
represent the Poissonian error of the detection counts. Solid line represents the obtained $n(m)$ in this paper (see \S4.1). \label{bglf}}

\end{figure}

As an example, $n(m)$ values are shown in Fig.~\ref{bglf} for the three slopes
$\gamma=0.30$, $\gamma=0.34$, and $\gamma=0.39$ overplotted on the resolved
point sources in region ``Blank 3''. It can be seen that the three values of
$\gamma$ are in principle compatible with the resolved galaxy distribution
in this region.

The variance $\sigma_{\rm BG}^{2}$ can be now deduced as the second moment of
$n(m)$, and $\sigma_{\rm IGC}^2$ can be obtained from equation~\ref{sigmaIGC}. In
Table~\ref{igc}, and for each value of $\gamma$, the results associated with
each galaxy for $\sigma_{\rm BG}^{2}$ are listed in column 4. The error bars of $\sigma_{\rm BG}^{2}$ have been computed taking into account only
the $n(m)$ fitting uncertainty. Note that in \citet{MA02} the error bar of
$\sigma_{\rm BG}^{2}$ was determined considering possible variations in the
slope of the faint end of the $n(m)$ (from $\gamma=0.30$ to $\gamma=0.40$),
but in the present study, these possible variations are being considered
explicitly by studying the cases $\gamma=0.30$, $0.34$, and $0.39$ separately. 

\subsection{Intergalactic globular clusters}

Equation \ref{sigmaIGC} can now be used to derive $\sigma_{\rm IGC}^2$ for each
region. The results are given in column 5 of Table \ref{igc}. The number
population of IGC, $N_{\rm IGC}$, can be estimated from $\sigma_{\rm IGC}^2$ using
the $\sigma^{2}_{\rm IGC}/N_{\rm IGC}$ ratios listed column 6. These ratios were computed in \citet{MA02}. The resulting $N_{\rm IGC}$ is listed in column 7. Note that $N_{\rm IGC}$ is expresed in $IGCs/(\arcsec)^2$, while the rest of columns are expresed in $pixels$. The $\sigma_{\rm IGC}^2$ uncertainties have been computed adding in quadrature the uncertainties of $P_0$ and $\sigma_{\rm BG}^2$. On the other hand, the uncertainties of the ratio $\sigma^{2}_{\rm IGC}/N_{\rm IGC}$ have been neglected. The reason is that
$\varphi$ is assumed to be universal (we assume the same $\varphi$ parameters
adopted in Mar\'\i n-Franch \& Aparicio 2002, 
$\sigma=1.40$ and $m_R^0=27.42$). In this sense, the
uncertainty in the $\varphi$ parameters would introduce a rescaling factor
common to all the results, but the relative errors would not change and the
proofs of IGCs detection would remain valid. The origin of the galaxy-to-galaxy variations in the ratio $\sigma^{2}_{\rm IGC}/N_{\rm IGC}$ relies on the differences of the photometric calibration constants.

\begin{figure} 
\plotone{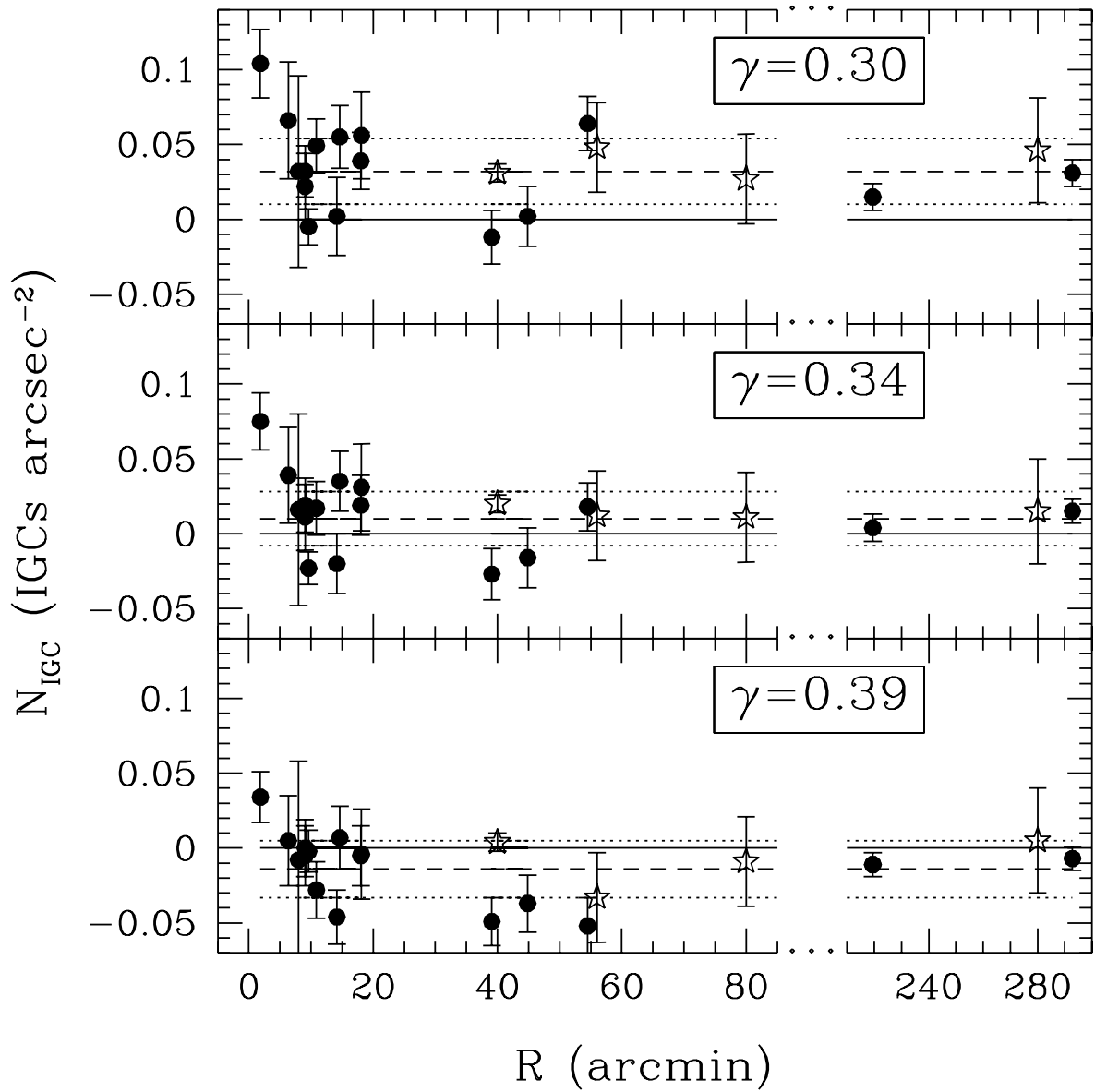}
\caption{The IGC population in Coma plotted for three possible values of the $n(m)$ slope, $\gamma$. Horizontal axis represents the distance to the Coma galaxy cluster center, and vertical axis represents the IGCs surface number density. Full lines indicate the zero density level. Dashed lines represent the mean for each case while dotted lines show the $\pm1\sigma$ of the mean. Full circles represent the outskirts of the 16  galaxies studied, while open stars represent the 
four control fields. See text for details. \label{gamma}}

\end{figure}

The values obtained for $N_{\rm IGC}$ for the field regions associated 
with the 16
galaxies studied in \cite{MA02} (dots) and for four additional intergalactic
field regions (stars) are plotted in Fig.~\ref{gamma} versus the distance $R$
to the Coma galaxy cluster center. Each panel corresponds to a value of
$\gamma$. Regions associated with galaxies were selected in \cite{MA02} with the
aim of sampling the background. They are beyond the GCS of each galaxy and
they consequently sample the intergalactic field population. The four
intergalactic ``blank'' regions are far from any galaxy. The fact that the
average of $N_{\rm IGC}$ for these regions is the same as for the regions
associated with galaxies confirms that the field is really being sampled in
both cases.

The most significant characteristic of the plots in Fig.~\ref{gamma} is the flat
distributions of values. The only point perhaps departing from this trend is
that associated with the central galaxy, NGC 4874. Furthermore, it can be
seen that $N_{\rm IGC}$ is negative in almost all cases for $\gamma=0.39$,
indicating that this value is unlikely. On the other hand, if $\gamma=0.30$,
then $N_{\rm IGC}$ is clearly a positive number, implying that an IGC population
should exist. This population is also detected if $\gamma=0.34$, but less
clearly. It should be noted that, if $\gamma$ = 0.30--0.34, $N_{\rm IGC}$
does not correlate with $R$, and hence with the environment, except for the
fact that the density associated with the central galaxy is higher. Rather, the
IGC population would be uniformly spread all over Coma. We will discuss
this issue further in Section~4.

For each value of $\gamma$, we have computed the average $\langle N_{\rm IGC}
\rangle$ in Coma
using all  but the central points. The results are listed in Table~\ref{meanIGC} and have been plotted in Fig.~\ref{gamma} by dashed
lines. Dotted lines show the 1 $\sigma_{\rm points}$ intervals ($\sigma_{\rm points}$
is the root mean square of the data point distribution). The estimated errors
of the mean have also been computed using the usual formula
\begin{equation}
\sigma_{\langle N_{\rm IGC}\rangle}=\frac{\sigma_{\rm points}}{\sqrt{N-1}},
\end{equation}
and are the errors quoted in Table~\ref{meanIGC}.

\section{Discussion}

\subsection{The faint end of the differential galaxy number counts, $n(m)$}

The extreme cases  considered above for the $n(m)$ slope, $\gamma=0.30$ and
$\gamma=0.39$, characterize two distinct scenarios. We have seen that the
latter produces negative $N_{\rm IGC}$ all over Coma, which indicates that $n(m)$
must be shallower.

Let us consider now in more detail the possibility $\gamma=0.30$. In this case,
Fig.~\ref{gamma} shows that IGCs would exist and would be spread all over
Coma. Furthermore, no relation with the distance to the galaxy cluster center is
apparent. However, all the GC formation scenarios mentioned in \S \ref{intro}
imply that $N_{\rm IGC}$ should follow the galaxy cluster mass distribution. Assuming this
and in order to test our results, we will consider three different galaxy cluster
simplistic mass distributions, $\rho(R)$, and analyze the expected $N_{\rm IGC}$
distribution for each model.

\begin{figure} 
\plotone{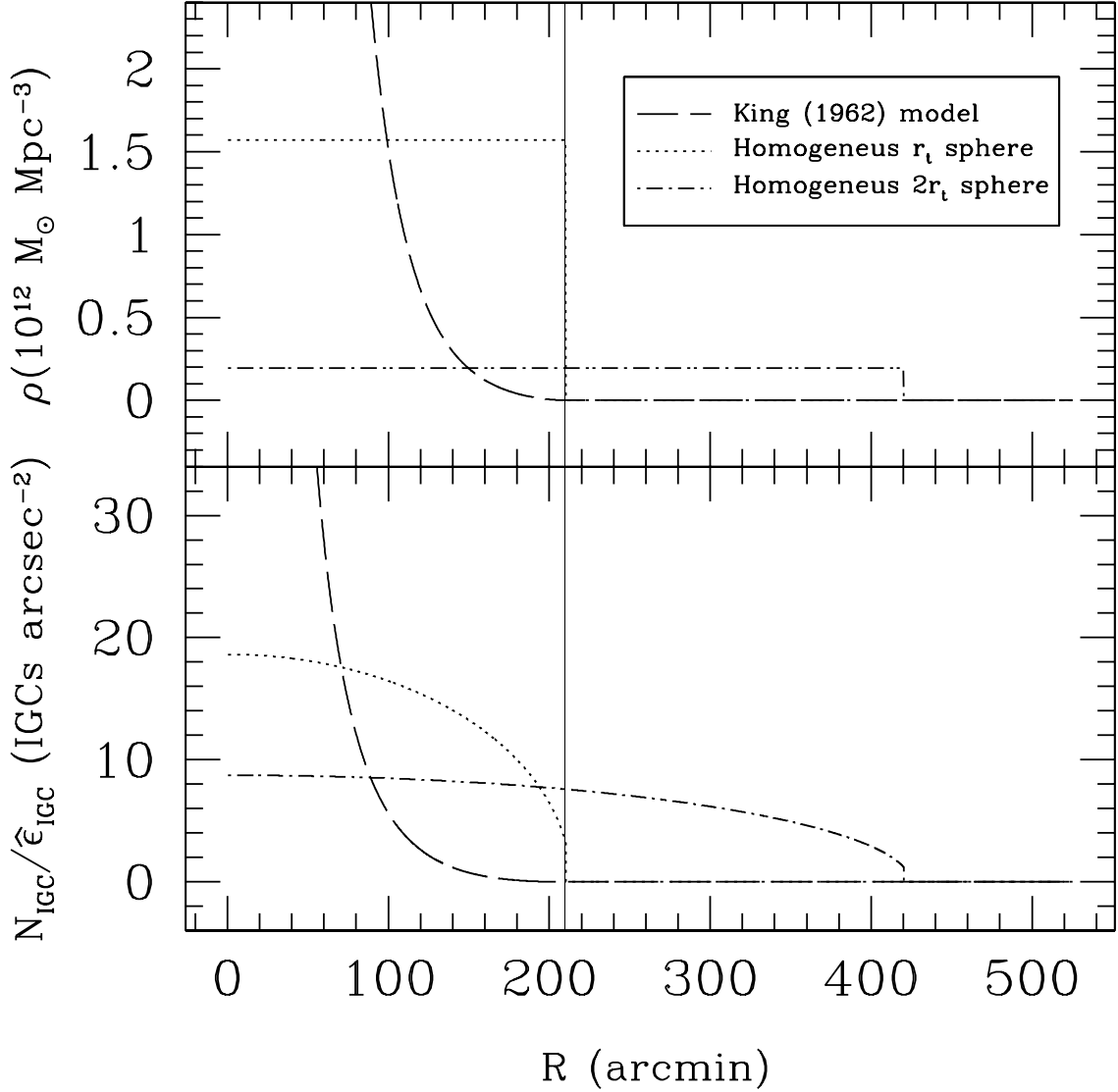}
\caption{Upper panel: The total mass density ($\rho$) is plotted as a
function of the distance to the galaxy cluster center ($R$) for three different galaxy cluster mass
density distributions; \citet{K62} model (long dash), homogeneous sphere of
radius equal to the Coma radius (dots) and homogeneous sphere of radius twice
the Coma radius (short dash). Lower panel: IGC surface number density
normalized to the globular cluster formation efficiency
($N_{\rm IGC}/\hat\epsilon_{\rm IGC}$) for each of the former mass distributions. See
text for details. \label{massdis}}

\end{figure}

\citet{M94} has given for Coma a cut-off radius $r_t=3.5\arcdeg$, a core
radius $r_c=0.24\arcdeg$ (i.e. $r_t/r_c=14.2$) and, using $H_0=72\pm8$ km
s$^{-1}$ Mpc$^{-1}$ \citep{F01} for the Hubble parameter, a total mass
$1.4\times10^{15} M_{\odot}$. This value for the Hubble parameter corresponds
to a distance of 97.2 Mpc and to $r_t=5.9$ Mpc and $r_c=0.4$ Mpc. In this context, the three considered galaxy cluster mass distribution models correspond to a King model, a
homogeneous sphere of radius equal to the cut-off radius of Coma ($r_t=210 \arcmin$) and a homogeneous sphere of radius equal to twice the cut-off radius
of Coma. All them are plotted in the upper panel of Figure \ref{massdis}. All
the mass distributions are normalized to a total galaxy cluster mass of $1.4\times10^{15}
M_{\odot}$ \citep{M94}. The vertical solid line represents the Coma cut-off radius,
$r_t$.

Once a mass distribution model is assumed, the expected $N_{\rm IGC}$ can be
deduced projecting the galaxy cluster spatial mass distribution and taking into account a
GC formation efficiency. The latter is a measure of the fraction of mass
converted into GCs. \citet{M99} defined the GC formation efficiency in galaxies
as
\begin{equation}
\epsilon_{\rm GC}=\frac{\rho_{GC}}{\rho_{\rm gas}+\rho_{\rm stars}},
\end{equation}
where $\rho_{\rm GC}$, $\rho_{\rm gas}$, and $\rho_{\rm 
stars}$ are the mass densities of GCs,
gas, and stars in the galaxy, respectively. \citet{M99} found this
definition of $\epsilon_{GC}$ to be universal from galaxy to galaxy and equal to $0.0026 \pm 0.0005$.

We will use here a similar definition for the intergalactic medium that,
for convenience, we define  in terms of surface density:
\begin{equation}
\hat\epsilon_{\rm IGC}=\frac{\Sigma_{\rm IGC}}{\Sigma_{\rm tot}}.
\end{equation}
where $\Sigma_{\rm IGC}$ is the surface mass density of IGCs and $\Sigma_{\rm tot}$ is
the baryonic surface mass density of the galaxy cluster obtained projecting $\rho(R)$. 

The relation between the baryonic surface mass density and the IGC surface
number density is given by
\begin{equation}
N_{\rm IGC}=\frac{\hat\epsilon_{\rm IGC}\Sigma_{\rm tot}}{\langle m\rangle_{\rm GC}},
\end{equation}
where $\langle m\rangle_{\rm GC}$ is the mean mass of a GC. Assuming $\langle m\rangle_{\rm GC}=2.4\times10^5$
\citep{M99}, the lower panel of Fig.~\ref{massdis} shows
$N_{\rm IGC}/\hat\epsilon_{\rm IGC}$ as a function of $R$ for the three mass
distribution models plotted in the upper panel. This figure shows that the
baryonic mass distribution model required to produce an approximately
constant, non-null, $N_{\rm IGC}$ distribution along Coma (up to $R\simeq 200 \arcmin$, at least) should be similar to a very extended homogeneous
sphere. The unrealistic nature of this model indicates that a value as low as
$\gamma=0.30$ for the slope of the faint end of $n(m)$ is unlikely. Moreover,
it suggests that the IGC surface density, $N_{\rm IGC}$, is actually zero all over
Coma, with the probable exception of the surroundings of the central galaxy.
The value of $\gamma$ that reduces the mean $\langle N_{\rm IGC}\rangle$ to  zero (excluding the
central value) can be calculated from interpolation in the data listed in
Table~\ref{meanIGC} and turns out to be $\gamma=0.36\pm0.01$.

The former is the slope of $n(m)$ beyond the detection limit of $R=23.5$
reached by \citet{T88} and \citet{SH93} and  is valid at least down to the
$R$ magnitude significantly contributing to the SBF. This magnitude can be
evaluated by  considering that in Table~\ref{igc} the error bars associated 
with
$\sigma_{\rm BG}^2$ are, on average, about 5\%. If $\sigma_{\rm BG}^2$ is computed
for different limiting magnitudes, it is found that the variance produced by
background galaxies fainter than $m_R=26.5$ is lower than this
value of  5\%. The
signal produced by fainter galaxies remains below the error bars. In
conclusion, $m_R=26.5$ can be assumed the faintest limit of our
determination, so that the slope of the faint end of $n(m)$ is
$\gamma=0.36\pm0.01$ (this slope is valid down to $m_R=26.5$ at least).

For comparison, it is interesting to note that \citet{W96} have measured the
slope of $n(m)$ for deeper $V$ and $I$ magnitudes from the Hubble Deep
Field. They obtain $\gamma=0.36$ for $V$ and $\gamma=0.31$ for $I$ for the magnitude interval [23,26], and $\gamma=0.17$ for $V$ and $\gamma=0.18$ for $I$ for the magnitude interval [26,29].

\subsection{Intergalactic globular clusters}

In \S4.1 we  concluded that the most likely value for the general IGC
density in Coma is zero, and that this is accomplished with a slope
$\gamma=0.36\pm 0.01$ for the faint end of the background differential galaxy number counts, $n(m)$. We will now further discuss the implications of these
results on the IGC population and the overall baryonic mass distribution in
Coma.

\begin{figure} 
\plotone{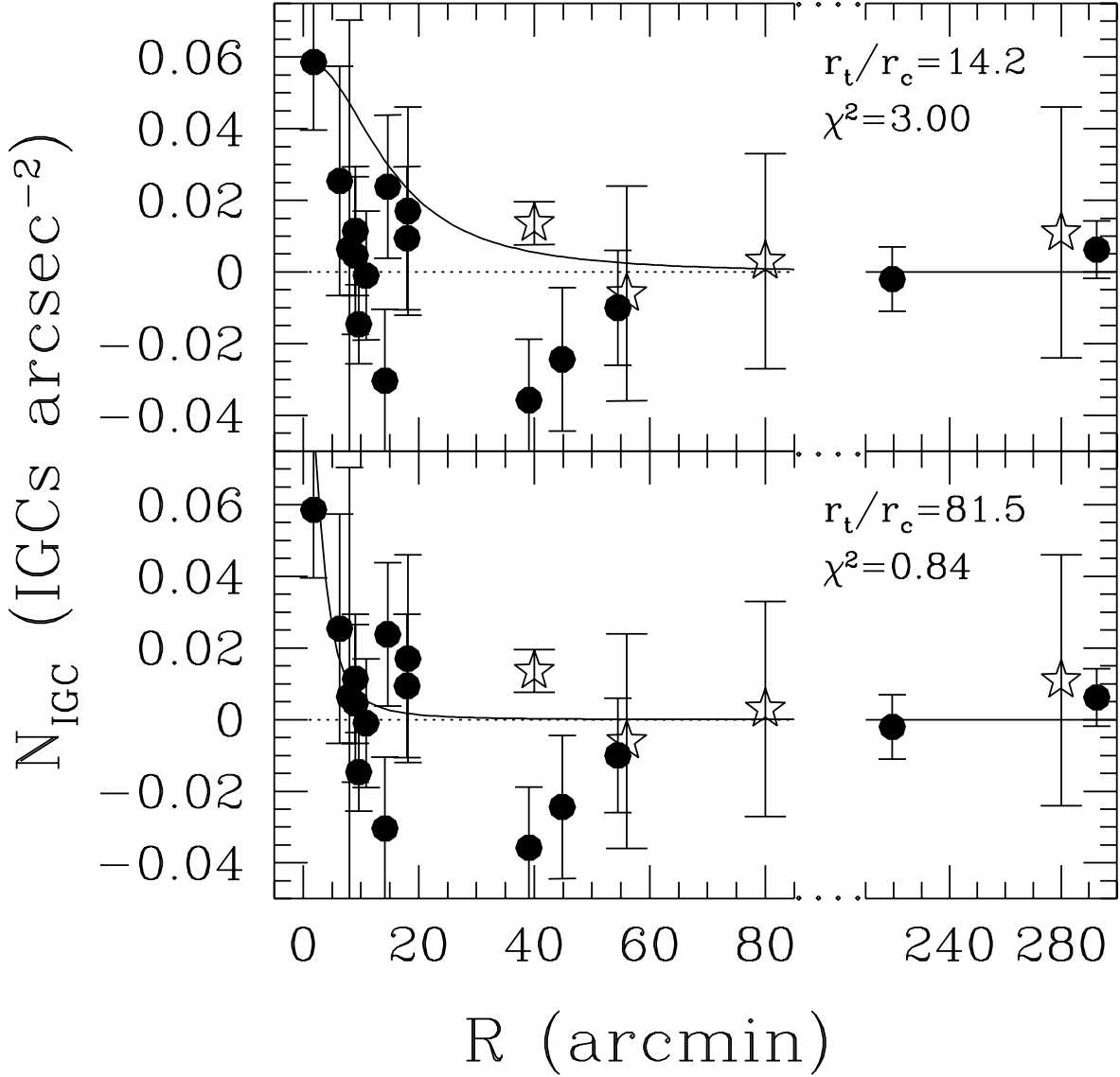}
\caption{$N_{\rm IGC}$ interpolated for the resulting case $\gamma=0.36$. Upper panel: The theoretical expectation from the \citet{K62} model has been fitted to our central $N_{\rm IGC}$ result considering the \citet{M94} core radius (solid line); Lower panel: $N_{\rm IGC}$ results have been fitted with a \citet{K62} model considering $r_t/r_c$ and $\hat\epsilon_{\rm IGC}$ as free parameters. The best fit is plotted with a solid line. \label{massdis2}}

\end{figure}

In Fig.~\ref{massdis2}, $N_{\rm IGC}$ obtained with $\gamma=0.36$ is plotted as a
function of $R$, the distance to the Coma center. The points have
been computed as an interpolation between the $\gamma=0.34$ and $\gamma=0.39$
results listed in Table \ref{igc}. It can be seen that the mean 
$\langle N_{\rm IGC}\rangle$
is zero, as required. Only NGC~4874 seems to be surrounded by a significant
number of IGCs. A King model, such as that plotted in Fig.~\ref{massdis},
 can now be
compared with the IGC distribution. The efficiency $\hat\epsilon_{\rm IGC}$ can be
calibrated from the central IGC density $N_{\rm IGC}=0.059$ 
$(\arcsec)^{-2}$ and is $\hat\epsilon_{\rm IGC}=7.01 \times 10^{-5}$. Incidentally, this value is two orders of magnitude smaller than that obtained by \citet{M99} for GCs belonging to galaxies.

The former $\hat\epsilon_{\rm IGC}$ and the concentration parameter
$r_t/r_c=14.2$ derived by \citet{M94} for the intergalactic matter in Coma
associated with the X-ray radiation can be used to plot the expected
distribution of IGCs. The result is shown in Fig.~\ref{massdis2}, upper panel
(solid line). This would be the distribution of IGC in Coma if: (i) the X-ray
emission properly samples the baryonic intergalactic mass distribution in
Coma; (ii) the GC excess observed in the central region is caused by a
genuine IGC population bound by the general Coma potential well, and (iii) if
the efficiency $\hat\epsilon$ is constant all over Coma. This distribution
lies beyond the error bars of several single $N_{\rm IGC}$ values. A higher
concentration parameter is required to properly fit the $N_{\rm IGC}$ point
distribution.  The best fit to a King distribution produces $r_t/r_c=81.5$
and $\hat\epsilon_{\rm IGC}=9.5\times 10^{-5}$. The result has
been plotted in the lower panel of Fig.~\ref{massdis2}.

The reduced chi-square for the best fit results $\chi^2_{\nu}=0.84$ which
indicates that the fit is good enough. On the other hand, $\chi^2_{\nu}=3.0$ for
the $r_t/r_c=14.2$ king model fit. For $\nu=19$ (as in our case), the
integral probability of the $\chi^2_{\nu}$ distribution is 0.001 for
$\chi^2_{\nu} \geq 2.31$, showing that the $r_t/r_c=14.2$ king model fit can be
rejected with a high confidence level.

The concentration parameter $r_t/r_c=81.5$, implies a core radius of
$r_c=154\arcsec$. This is of the order of, or smaller than, the size of the
Coma central galaxy NGC 4874. In fact, in \citet{MA02}, the spatial structure
of the NGC 4874 GCS was studied out to a distance of $161.4\arcsec$ from the
galaxy center, and its edge was probably not reached. This suggests that the
obtained $N_{\rm IGC}$ excess around the center of Coma is produced by the
outermost extension of the NGC 4874 primitive GCS rather than being a trace
of an IGC system. In other words, the spatial distribution of $N_{\rm IGC}$
shown in Fig. \ref{massdis2} is probably the map of the NGC~4874 GCS.

This result suggests that an IGC population does not exist in Coma. This
points toward GCs having been formed only, or almost only, from protogalactic
clouds. None of them, or perhaps very few, could have formed in isolated
regions. This is in good agreement with the {in situ} formation scenario for
the GCSs \citep{B97,F97,H98,M99}. If the distribution of IGCs followed the
distribution of dark matter in galaxy clusters \citep{M87,W93}, it would be
reasonable to expect that $N_{\rm IGC}$ would show a spatially extended
gradient across de galaxy cluster. This not being the case, it seems
inappropriate to support the relationship between IGC and dark matter
distributions, although it is true that the relationship could still exist
but not be strong enough to be detected.

On the other hand, a number of authors have argued that accretion of IGCs
might influence the formation and evolution of galaxies
\citep{M87,W87,W95,C01}. Our results indicate that accretion of IGCs is not a
significant effect in galaxy formation and evolutionary processes in the Coma
galaxies, since our conclusion is that IGCs do not exist in Coma.

\citet{GM01} found the existence of three subgroups of galaxies in Coma, one
of them associated with the cD galaxy NGC 4874 and the other two with NGC~4889
and NGC~4839. They conclude that the non-stationarity of the dynamical
processes at work in the Coma core is due to the merging of small-scale
groups of galaxies. In this context, each subgroup formed separately and then
the merger between the different groups took place. In \citet{MA02}, five
galaxies belonging to the \citet{GM01} subgroup 1 (the NGC 4874 group) were
analyzed showing that a relation between $S_N$ and the distance to the
central galaxy NGC 4874 was compatible with the data. The possibility of
accretion of IGCs being responsible for this relation was discussed. In the
light of our new results and owing to the absence of IGCs found here, this
possibility should be also rejected.

\section{Conclusions}

The existence of an IGC population in the Coma galaxy cluster has been tested using 
the SBF. The measured SBF signal is mainly produced by faint galaxies and the IGC population (if it exists). For this reason, the  $N_{\rm IGC}$ results
reached and the $n(m)$ faint end are connected. As we do not have a precise enough measurement of $n(m)$ in the Coma galaxy cluster, $N_{\rm IGC}$ has been obtained considering three different values for the $n(m)$ slope ($\gamma=0.30$, $0.34$, and $0.39$) and the results have been analyzed separately. The main conclusions of this study are summarized here:

\begin{itemize}
\item
For the three  $\gamma$ values studied, the   results obtained for $N_{\rm IGC}$ have been shown not to correlate with the distance to the central galaxy NGC 4874, and hence with the environment. Comparing the $N_{\rm IGC}$ results obtained for each value of $\gamma$ with the different Coma galaxy cluster mass distribution model predictions, we suggest that $N_{\rm IGC}$ must be zero all over Coma, with the probably exception of the surroundings of the central galaxy. The most likely $n(m)$ slope that makes $\langle N_{\rm IGC}\rangle=0$ is $\gamma=0.36\pm0.01$. This slope has been shown to be valid up to $m_R=26.5$. For comparison, previous results for the $n(m)$ slope in the $R$ filter range from $\gamma=0.39$ \citep{T88} to $\gamma=$ 0.31--0.34 \citep{SH93}. This slope has also been measured in the Hubble Deep Field North in the magnitude interval [23,26], resulting in $\gamma=0.36$ for the $V$ filter and $\gamma=0.31$ for the $I$ filter \citep{W96}. Our new $\gamma$ determination falls between previous measurements. 

\item
Our results seem to be in good agreement with the {in situ} formation for the GCSs \citep{B97,F97,H98,M99}. The former conclusions point toward GCs having been formed only, or almost only, from protogalactic clouds. None of them, or perhaps very few, could have formed in isolated regions. 

\item
\citet{M87} and \citet{W93} argued that there might exist a relationship between IGC and dark matter distributions. In the light of our results it seems inappropriate to support this idea, although it is true that the relationship could still exist but not be strong enough to be detected.

\item
Finally, a number of authors have argued that accretion of IGCs might influence the formation and evolution of galaxies \citep{M87,W87,W95,C01}. Our results suggest that accretion of IGCs might not be a significant effect in galaxy formation and the evolutionary processes in the Coma galaxies since our conclusion is that IGCs do not exist in Coma.

\end{itemize}

\acknowledgments
 
This article is based on observations made with the 2.5 m Isaac Newton Telescope operated on the island of La Palma by the ING in the Spanish Observatorio del Roque de Los Muchachos. This research has been supported by the Instituto de Astrof\'\i sica de Canarias (grant P3/94), the DGESIC of the Kingdom of Spain (grant PI97-1438-C02-01), and the DGUI of the autonomous government of the Canary Islands (grant PI1999/008).

\clearpage

\clearpage

\clearpage

\clearpage

\clearpage

\clearpage
\begin{deluxetable}{lr}
\tabletypesize{\scriptsize}
\tablecaption{Notation summary. \label{nota}}
\tablewidth{0pt}
\tablehead{}

\startdata

Globular Cluster                     & GC        \\
Intergalactic Globular Cluster       & IGC       \\ 
Globular Cluster System              & GCS       \\
Surface Brightness Fluctuations      & SBF       \\
Differential Galaxy Number Counts    & $n(m)$    \\
Differential Galaxy Number Counts Slope     & $\gamma$  \\
Globular Cluster Luminosity Function & $\varphi$ \\

\enddata

\end{deluxetable}

\clearpage

\begin{deluxetable}{lccccc|r}
\tabletypesize{\scriptsize}
\tablecaption{IGCs in the environments of the 16 Coma galaxies\tablenotemark{(a)}     .\label{igc}}

\tablewidth{0pt}
\tablehead{
\colhead{Galaxy} & 
\colhead{$R$} & 
\colhead{$P_0$} & 
\colhead{$\sigma_{\rm BG}^{2}$} & 
\colhead{$\sigma_{\rm IGC}^2$} & 
\colhead{$\sigma_{\rm IGC}^2/N_{\rm IGC}$\tablenotemark{(b)}} & 
\colhead{$N_{\rm IGC}$} \\
& 
($\arcmin$) & 
$10^{-4} \times \left(\frac{e^-}{\rm s \times pix}\right)^2$  &
$10^{-4} \times \left(\frac{e^-}{\rm s \times pix}\right)^2$  &
$10^{-4} \times \left(\frac{e^-}{\rm s \times pix}\right)^2$  &
&
${\rm IGCs}/(\arcsec)^2$  
}

\startdata
\multicolumn{7}{c}{\bf{\small $\gamma=0.30$ }} \\
\noalign{\smallskip}
\hline
\hline

NGC 4874             & 0       & 31.8 $\pm$ 1.2 & 22.6  $\pm$ 1.2  &  7.7  $\pm$  1.7 & 0.067 & 0.104 $\pm$ 0.023  \\
NGC 4889             & 8.05    & 26.2 $\pm$ 2.1 & 19.8  $\pm$ 1.4  &  4.9  $\pm$  2.9 & 0.067 & 0.066 $\pm$ 0.039  \\
IC 4012              & 10.68   & 18.0 $\pm$ 1.1 & 14.9  $\pm$ 1.2  &  1.6  $\pm$  1.6 & 0.067 & 0.022 $\pm$ 0.022  \\
IC 4021              & 10.88   & 16.4 $\pm$ 1.0 & 12.5  $\pm$ 0.8  &  2.4  $\pm$  1.3 & 0.067 & 0.032 $\pm$ 0.017  \\
IC 4026              & 12.62   & 17.2 $\pm$ 1.1 & 12.0  $\pm$ 0.9  &  3.7  $\pm$  1.4 & 0.067 & 0.049 $\pm$ 0.018  \\
IC 4041              & 16.38   & 21.2 $\pm$ 1.1 & 15.6  $\pm$ 1.1  &  4.1  $\pm$  1.6 & 0.067 & 0.055 $\pm$ 0.021  \\
IC 4045              & 19.85   & 19.7 $\pm$ 1.1 & 15.3  $\pm$ 0.9  &  2.9  $\pm$  1.4 & 0.067 & 0.039 $\pm$ 0.019  \\
IC 4051              & 19.95   & 26.1 $\pm$ 2.1 & 20.4  $\pm$ 0.8  &  4.2  $\pm$  2.2 & 0.067 & 0.056 $\pm$ 0.029  \\
IC 3976              & 6.78    & 11.5 $\pm$ 3.0 &  8.4  $\pm$ 1.3  &  1.6  $\pm$  3.2 & 0.045 & 0.032 $\pm$ 0.064  \\
IC 3959              & 12.67   & 15.2 $\pm$ 1.0 & 13.6  $\pm$ 0.8  &  0.1  $\pm$  1.3 & 0.059 & 0.002 $\pm$ 0.026  \\
MCG +5 $-$31 $-$063  & 8.93    & 13.5 $\pm$ 0.6 & 12.3  $\pm$ 0.3  &--0.3  $\pm$  0.7 & 0.059 &--0.005$\pm$ 0.012  \\
NGC 4839             & 43.12   & 10.3 $\pm$ 1.0 &  8.7  $\pm$ 0.1  &  0.1  $\pm$  1.0 & 0.046 & 0.002 $\pm$ 0.020  \\
NGC 4840             & 37.43   & 9.5  $\pm$ 0.8 &  8.6  $\pm$ 0.3  &--0.6  $\pm$  0.9 & 0.046 &--0.012$\pm$ 0.018  \\
NGC 4816             & 52.60   & 16.5 $\pm$ 1.1 & 10.8  $\pm$ 0.4  &  4.2  $\pm$  1.2 & 0.060 & 0.064 $\pm$ 0.018  \\
NGC 4673             & 217.07  & 15.5 $\pm$ 0.6 & 12.9  $\pm$ 0.4  &  1.1  $\pm$  0.7 & 0.067 & 0.015 $\pm$ 0.009  \\
IC 3651              & 290.32  & 16.7 $\pm$ 0.6 & 12.9  $\pm$ 0.3  &  2.3  $\pm$  0.7 & 0.067 & 0.031 $\pm$ 0.009  \\
Blank 1             & 40      & 10.3 $\pm$ 0.3 & 6.7   $\pm$ 0.4  &  2.1  $\pm$  0.5 & 0.060 & 0.031 $\pm$ 0.007  \\
Blank 2             & 56      & 14.5 $\pm$ 2.0 & 9.8   $\pm$ 0.3  &  3.2  $\pm$  2.0 & 0.060 & 0.048 $\pm$ 0.030  \\
Blank 3             & 80      & 11.0 $\pm$ 2.0 & 7.7   $\pm$ 0.3  &  1.8  $\pm$  2.0 & 0.060 & 0.027 $\pm$ 0.030  \\
Blank 4             & 280     & 16.0 $\pm$ 3.0 & 11.4  $\pm$ 0.5  &  3.1  $\pm$  3.0 & 0.060 & 0.046 $\pm$ 0.044  \\

\hline
\noalign{\smallskip}
\multicolumn{7}{c}{\bf{\small $\gamma=0.34$ }} \\
\noalign{\smallskip}
\hline
\hline

NGC 4874            & 0       & 31.8 $\pm$ 1.2 & 24.7  $\pm$ 0.8 & 5.6  $\pm$  1.4 & 0.067 & 0.075   $\pm$ 0.019  \\
NGC 4889            & 8.05    & 26.2 $\pm$ 2.1 & 21.8  $\pm$ 1.2 & 2.9  $\pm$  2.4 & 0.067 & 0.039   $\pm$ 0.032  \\
IC 4012             & 10.68   & 18.0 $\pm$ 1.1 & 15.7  $\pm$ 1.2 & 0.8  $\pm$  1.6 & 0.067 & 0.011   $\pm$ 0.022  \\
IC 4021             & 10.88   & 16.4 $\pm$ 1.0 & 13.5  $\pm$ 0.9 & 1.4  $\pm$  1.3 & 0.067 & 0.019   $\pm$ 0.018  \\
IC 4026             & 12.62   & 17.2 $\pm$ 1.1 & 14.4  $\pm$ 0.9 & 1.3  $\pm$  1.4 & 0.067 & 0.017   $\pm$ 0.018  \\
IC 4041             & 16.38   & 21.2 $\pm$ 1.1 & 17.1  $\pm$ 1.0 & 2.6  $\pm$  1.5 & 0.067 & 0.035   $\pm$ 0.020  \\
IC 4045             & 19.85   & 19.7 $\pm$ 1.1 & 16.8  $\pm$ 1.0 & 1.4  $\pm$  1.5 & 0.067 & 0.019   $\pm$ 0.020  \\
IC 4051             & 19.95   & 26.1 $\pm$ 2.1 & 22.3  $\pm$ 0.8 & 2.3  $\pm$  2.2 & 0.067 & 0.031   $\pm$ 0.029  \\
IC 3976             & 6.78    & 11.5 $\pm$ 3.0 &  9.2  $\pm$ 1.3 & 0.8  $\pm$  3.2 & 0.045 & 0.016   $\pm$ 0.064  \\
IC 3959             & 12.67   & 15.2 $\pm$ 1.0 & 15.0  $\pm$ 0.8 &--1.3 $\pm$  1.3 & 0.059 &--0.020  $\pm$ 0.020  \\
MCG +5 $-$31 $-$063 & 8.93    & 13.5 $\pm$ 0.6 & 13.4  $\pm$ 0.3 &--1.5 $\pm$  0.7 & 0.059 &--0.023  $\pm$ 0.011  \\
NGC 4839            & 43.12   & 10.3 $\pm$ 1.0 &  9.6  $\pm$ 0.1 &--0.8 $\pm$  1.0 & 0.046 &--0.016  $\pm$ 0.020  \\
NGC 4840            & 37.43   & 9.5  $\pm$ 0.8 &  9.4  $\pm$ 0.3 &--1.4 $\pm$  0.9 & 0.046 &--0.027  $\pm$ 0.017  \\
NGC 4816            & 52.60   & 16.5 $\pm$ 1.1 & 13.8  $\pm$ 0.3 & 1.2  $\pm$  1.1 & 0.060 & 0.018   $\pm$ 0.016  \\
NGC 4673            & 217.07  & 15.5 $\pm$ 0.6 & 13.7  $\pm$ 0.3 & 0.3  $\pm$  0.7 & 0.067 & 0.004   $\pm$ 0.009  \\
IC 3651             & 290.32  & 16.7 $\pm$ 0.6 & 14.1  $\pm$ 0.2 & 1.1  $\pm$  0.6 & 0.067 & 0.015   $\pm$ 0.008  \\
Blank 1            & 40      & 10.3 $\pm$ 0.3 & 7.4   $\pm$ 0.4 &  1.4 $\pm$  0.5 & 0.060 & 0.021   $\pm$ 0.007  \\
Blank 2            & 56      & 14.5 $\pm$ 2.0 & 12.2  $\pm$ 0.3 &  0.8 $\pm$  2.0 & 0.060 & 0.012   $\pm$ 0.030  \\
Blank 3            & 80      & 11.0 $\pm$ 2.0 & 8.7   $\pm$ 0.3 &  0.8 $\pm$  2.0 & 0.060 & 0.012   $\pm$ 0.030  \\
Blank 4            & 280     & 16.0 $\pm$ 3.0 & 12.5  $\pm$ 0.5 &  1.0 $\pm$  3.0 & 0.060 & 0.015   $\pm$ 0.045  \\
\hline
\noalign{\smallskip}
\multicolumn{7}{c}{\bf{\small $\gamma=0.39$ }} \\
\noalign{\smallskip}
\hline
\hline

NGC 4874            & 0       & 31.8 $\pm$ 1.2 & 27.7  $\pm$ 0.6  &  2.6  $\pm$  1.3  & 0.067 &  0.034   $\pm$ 0.017 \\
NGC 4889            & 8.05    & 26.2 $\pm$ 2.1 & 24.3  $\pm$ 1.2  &  0.4  $\pm$  2.4  & 0.067 &  0.005   $\pm$ 0.030 \\
IC 4012             & 10.68   & 18.0 $\pm$ 1.1 & 16.9  $\pm$ 1.2  &--0.4  $\pm$  1.6  & 0.067 &--0.005   $\pm$ 0.020 \\
IC 4021             & 10.88   & 16.4 $\pm$ 1.0 & 14.9  $\pm$ 1.1  &  0.0  $\pm$  1.5  & 0.067 &  0.000   $\pm$ 0.019 \\
IC 4026             & 12.62   & 17.2 $\pm$ 1.1 & 17.8  $\pm$ 0.9  &--2.1  $\pm$  1.4  & 0.067 &--0.028   $\pm$ 0.019 \\
IC 4041             & 16.38   & 21.2 $\pm$ 1.1 & 19.2  $\pm$ 1.0  &  0.5  $\pm$  1.5  & 0.067 &  0.007   $\pm$ 0.021 \\
IC 4045             & 19.85   & 19.7 $\pm$ 1.1 & 18.6  $\pm$ 1.2  &--0.4  $\pm$  1.6  & 0.067 &--0.005   $\pm$ 0.020 \\
IC 4051             & 19.95   & 26.1 $\pm$ 2.1 & 24.9  $\pm$ 0.9  &--0.3  $\pm$  2.3  & 0.067 &--0.004   $\pm$ 0.030 \\
IC 3976             & 6.78    & 11.5 $\pm$ 3.0 & 10.3  $\pm$ 1.4  &--0.4  $\pm$  3.3  & 0.045 &--0.008   $\pm$ 0.066 \\
IC 3959             & 12.67   & 15.2 $\pm$ 1.0 & 16.7  $\pm$ 0.7  &--3.0  $\pm$  1.2  & 0.059 &--0.046   $\pm$ 0.018 \\
MCG +5 $-$31 $-$063 & 8.93    & 13.5 $\pm$ 0.6 & 12.1  $\pm$ 0.3  &--0.1  $\pm$  0.7  & 0.059 &--0.002   $\pm$ 0.014 \\
NGC 4839            & 43.12   & 10.3 $\pm$ 1.0 & 10.7  $\pm$ 0.1  &--1.9  $\pm$  1.0  & 0.046 &--0.037   $\pm$ 0.019 \\
NGC 4840            & 37.43   & 9.5  $\pm$ 0.8 & 10.5  $\pm$ 0.2  &--2.5  $\pm$  0.8  & 0.046 &--0.049   $\pm$ 0.016 \\
NGC 4816            & 52.60   & 16.5 $\pm$ 1.1 & 18.5  $\pm$ 0.6  &--3.5  $\pm$  1.3  & 0.060 &--0.052   $\pm$ 0.019 \\
NGC 4673            & 217.07  & 15.5 $\pm$ 0.6 & 14.8  $\pm$ 0.2  &--0.8  $\pm$  0.6  & 0.067 &--0.011   $\pm$ 0.008 \\
IC 3651             & 290.32  & 16.7 $\pm$ 0.6 & 15.7  $\pm$ 0.2  &--0.5  $\pm$  0.6  & 0.067 &--0.007   $\pm$ 0.008 \\
Blank 1            & 40      & 10.3 $\pm$ 0.3 & 8.5   $\pm$ 0.4  &  0.3  $\pm$  0.5  & 0.060 &   0.004  $\pm$ 0.006 \\
Blank 2            & 56      & 14.5 $\pm$ 2.0 & 15.2  $\pm$ 0.3  &--2.2  $\pm$  2.0  & 0.060 &--0.033   $\pm$ 0.030 \\
Blank 3            & 80      & 11.0 $\pm$ 2.0 & 10.1  $\pm$ 0.3  &--0.6  $\pm$  2.0  & 0.060 &--0.009   $\pm$ 0.030 \\
Blank 4            & 280     & 16.0 $\pm$ 3.0 & 14.1  $\pm$ 0.5  &  0.4  $\pm$  3.0  & 0.060 &   0.006  $\pm$ 0.045 \\

\enddata

\tablenotetext{(a)}{Columns $P_0$, $\sigma_{\rm BG}^{2}$ and $\sigma_{\rm IGC}^2/N_{\rm IGC}$ have been extracted from the analysis done in \citet{MA02}. 1 $pix = 0.333\arcsec$.}
\tablenotetext{(b)}{Theoretical value based on the assumption that the $\varphi$ is universal with $\sigma=1.40$ and $m_R^0=27.42$. This value depends on the photometric calibration.}

\end{deluxetable}

\clearpage

\begin{deluxetable}{lcc}
\tabletypesize{\scriptsize}
\tablecaption{Mean surface number density of IGCs in Coma. \label{meanIGC}}
\tablewidth{0pt}
\tablehead{
\colhead{$\gamma$} &  \colhead{$\langle N_{\rm IGC}\rangle $} \\
 &  (${\rm IGCs}/(\arcsec)^2)$

}

\startdata

0.30 & 0.032    $\pm$  0.005  \\
0.34 & 0.010    $\pm$  0.004  \\
0.39 & --0.014  $\pm$  0.004  \\

\enddata

\end{deluxetable}


\clearpage
 
\begin{thebibliography}{}

\bibitem[Bahcall \& Soneira~(1981)]{BS81} Bahcall, J. N., \& Soneira, R. M. 1981, \apjs, 47, 357
\bibitem[Blakeslee et al.~(1997)]{B97} Blakeslee, J. P., Tonry, J. L., \& Metzger, M. R. 1997, \aj, 114, 482
\bibitem[C\^ot\'e et al.~(2001)]{C01} C\^ot\'e, P., McLaughlin, D. E., Hanes, D. A., Bridges, T. J., Geisler, D., Merritt, D., Hesser, J. E., Harris, G, L. H., \& Lee, M. G. 2001, \apj, 559, 828
\bibitem[Forbes, Brodie, \& Grillmair~(1997)]{F97} Forbes, D. A., Brodie, J. P., \& Grillmair, C. J. 1997, \aj, 113, 1652
\bibitem[Forte, Mart\' \i nez, \& Muzzio~(1982)]{F82} Forte, J. C., Mart\' \i nez, R. E., \& Muzzio, J. C. 1982, \aj, 87, 1465
\bibitem[Freedman et al.~(2001)]{F01} Freedman, W. L., Madore, B. F., Gibson, B. K., Ferrarese, L., Kelson, D. D., Sakai, S., Mould, J. R., Kennicutt, R. C. Jr., Ford, H. C., Graham, J. A., Huchra, J. P., Hughes, S. M. G., Illingworth, G. D., Macri, L. M., Stetson, P. B. 2001, \apj, 553, 47
\bibitem[Gurzadyan \& Mazure~(2001)]{GM01} Gurzadyan, V. G., \& Mazure, A. 2001, New Astronomy, 6, 43
\bibitem[Harris et al.~(1998)]{H98} Harris, W. E., Harris, G. L. H., \&  McLaughlin, D. E. 1998, \aj, 115, 1801
\bibitem[King~(1962)]{K62} King, I. R. 1962, \aj, 67, 471
\bibitem[Makino~(1994)]{M94} Makino, N., 1994, \pasj, 46, 139
\bibitem[Mar\' \i n-Franch \& Aparicio~(2002)]{MA02} Mar\' \i n-Franch, A., \& Aparicio, A. 2002, \apj, 568, 174
\bibitem[McLaughlin~(1999)]{M99} McLaughlin, D. E. 1999, \aj, 117, 2398
\bibitem[Muzzio~(1987)]{M87} Muzzio, J. C. 1987, \pasp, 99, 245
\bibitem[Steidel \& Hamilton~(1993)]{SH93} Steidel, C. C., \& Hamilton, D. 1993, \aj, 105, 2017
\bibitem[Tonry \& Schneider~(1988)]{TS88} Tonry, J. L., \& Schneider, D. P.  1988, \aj, 96, 807
\bibitem[Tyson~(1988)]{T88} Tyson, J. A. 1988, \aj, 96, 1
\bibitem[van den Bergh~(1956)]{B56} van den Bergh, S. 1956, \pasp, 68, 449
\bibitem[van den Bergh~(1958)]{B58} van den Bergh, S. 1958, The Observatory, 78, 85
\bibitem[West~(1993)]{W93} West, M. J. 1993, \mnras, 265, 755
\bibitem[West et al.~(1995)]{W95} West, M. J., C\^ot\'e, P., 
Jones, C., Forman, W., \& Marzke, R. O. 1995, \apj, 453, L77
\bibitem[White~(1987)]{W87} White, R. E. 1987, \mnras, 227, 185
\bibitem[Williams et al.~(1996)]{W96} Williams, R. E. et al., 1996, \aj, 112, 1335

\end{thebibliography}
\end{document}